\documentclass[prd,twocolumn,amsmath,amssymb,floatfix,superscriptaddress]{revtex4-1}

\usepackage{graphicx}
\usepackage{amssymb}
\usepackage{amsmath}
\usepackage{hyperref}

\usepackage{color}

\DeclareMathAlphabet\mathbfcal{OMS}{cmsy}{b}{n}
\definecolor{darkgreen}{cmyk}{0.85,0.2,1.00,0.2} 
\definecolor{purple}{cmyk}{0.5,1.0,0,0}

\def\barray{\begin{array}} 
\def\earray{\end{array}}
\def\be{\begin{equation}}
\def\ee{\end{equation}}
\def\ben{\begin{equation} \nonumber}
\def\een{\end{equation}}
\def\ban{\begin{eqnarray*}}
\def\ean{\end{eqnarray*}}
\def\ba{\begin{eqnarray}}
\def\ea{\end{eqnarray}}

\def\({\left(}
\def\){\right)}

\newcommand{\simgt}{\lower.5ex\hbox{$\; \buildrel > \over \sim \;$}}
\newcommand{\simlt}{\lower.5ex\hbox{$\; \buildrel < \over \sim \;$}}

\newcommand{\btheta}{{\boldsymbol{\theta}}}

\newcommand{\bl}{\boldsymbol{\ell}}
\newcommand{\el}{\ell}
\newcommand{\bL}{{\bf L}}

\newcommand{\fsky}{f_{\rm sky}}

\begin{document}

\title{Super-Sample CMB Lensing}
\author{Alessandro Manzotti}
\affiliation{Kavli Institute for Cosmological Physics, Department of Astronomy \& Astrophysics,  Enrico Fermi Institute, University of Chicago, Chicago, Illinois 60637, U.S.A.}
\author{Wayne Hu}
\affiliation{Kavli Institute for Cosmological Physics, Department of Astronomy \& Astrophysics,  Enrico Fermi Institute, University of Chicago, Chicago, Illinois 60637, U.S.A.}
\author{Aur{\'e}lien Benoit-L{\'e}vy}
\affiliation{Department of Physics and Astronomy, University College
London, London WC1E 6BT, United Kingdom}
%\date{}                                           % Activate to display a given date or no date
\begin{abstract}
Lensing of the CMB by modes that are larger than the size of the survey  dilates  intrinsic scales  in the temperature and polarization fields and coherently shifts their observed
power spectra with respect to the ensemble or all-sky mean.    The effect can be 
simply encapsulated  as a contribution to the power spectrum
covariance matrix in accordance with the lensing trispectrum or as an additional parameter,
the mean convergence in the field, for parameter estimation.  It should be included for upcoming surveys that precisely measure
acoustic polarization features deep into the damping tail at multipoles of $\ell \gtrsim 1500$ with less than $10\%$ of sky.   Its omission may lead to  seemingly conflicting values for the angular scale of the sound
horizon which may then provide erroneous cosmological parameters when compared
to  baryon acoustic oscillation measurements.
 \end{abstract}

\maketitle

\section{Introduction}

Gravitational lensing of the CMB is rapidly becoming both a useful tool for cosmology
and a necessary component to model carefully in the statistical analysis of temperature
and polarization anisotropy \cite{Das:2011ak,Das:2013zf,Keisler:2011aw,vanEngelen:2012va,Hanson:2013hsb,Ade:2013tya,Ade:2013tyw}.  In particular, as first detections
of CMB lensing in the $E$ and $B$ mode polarization power spectra become high 
significance measurements, it will become important to model non-Gaussian correlations
induced by lensing when characterizing their statistical errors  \cite{Smith06,BenoitLevy:2012va}.

For surveys that cover a large fraction of sky, the main contributions to non-Gaussian
errors come from the fact that the sample variance of the lensing power and the unlensed
CMB fields correlate power in the observed  spectra across a wide range of
multipoles \cite{BenoitLevy:2012va}.   However the next generation of surveys will
focus on deep polarization sensitive measurements on small patches of sky.  Here 
lensing by modes with wavelengths larger than the survey   modify the observed
power
spectrum in the sub-survey modes.   

 Even a small amount of super-sample lensing can 
produce a significant effect, since all sub-survey band power measurements covary.
In particular, lensing shifts
 the angular scale of the well-measured CMB acoustic peaks out as far in multipole
moment that they can be measured.
We call this effect super-sample covariance (SSC) following Ref.~\cite{Takada:2013wfa}
for the same effect in large-scale structure \cite{Hamilton:2005dx} that similarly
shifts the scale of baryon acoustic oscillations \cite{Li:2014sga}.
 Indeed it is also this same modulation of small-scale 
anisotropy modes by long-wavelength lensing modes that enables lensing reconstruction techniques and
squeezed bispectrum measurements of integrated-Sachs-Wolfe effect correlation with
lensing
\cite{Hu:2001tn,Okamoto:2002ik,Hanson:2010rp,Lewis:2011fk}.

In this paper, we study the effect of lensing-induced SSC on CMB power spectrum measurements for small patches of sky.   In \S \ref{sec:SSC} we derive the
band power covariance matrix for temperature and polarization measurements and
discuss the origin of SSC in the squeezed lensing trispectrum.   We give a simple criteria
for when SSC must be included in parameter estimation in \S \ref{sec:impact} and
discuss these results in \S \ref{sec:discussion}.

\section{Super-Sample Covariance}
\label{sec:SSC}

We introduce the SSC effect from lensing by super-sample modes for temperature
band power measurements in a finite sample in \S \ref{sec:temp}.   In \S \ref{sec:trispect}, 
we derive its form directly from the temperature trispectrum and in \S \ref{sec:pol}
we give the complete expressions for temperature and polarization band power covariance.

\subsection{Temperature Field}
\label{sec:temp}

We begin by considering the covariance of temperature power spectra estimators due to lensing of the CMB in a finite survey of angular area $A$.   A finite survey effectively measures the underlying temperature fluctuation field $T$ through
a mask or window 
\begin{equation}
T_W(\btheta) = T(\btheta) W(\btheta),
\end{equation}
where $W(\btheta)=1$ in the survey region and 0 otherwise.  As we shall review in more detail in \S \ref{sec:trispect}, 
in harmonic space  the effect of the window is to convolve the fields or correlate
band powers, destroying
independent information for multipoles separated by $\Delta \ell \lesssim \ell_W = 2\pi/\theta_W$
where $\theta_W \sim \sqrt{A}$ is the angular dimension of the survey.  

There are also physical effects that correlate band powers across $\Delta \ell \gg   2\pi/\theta_W$ from
gravitational lensing.   For simplicity, in this paper we shall work exclusively in this wide bin limit rather than deconvolve the window which would obscure the underlying physical
effects.
  These come about because two widely separated bands still jointly respond to the same 
modes in lensing potential.  
As such the covariance matrix for  power estimates $\hat C_{\ell_i}$ in bands of width  $\Delta \ell_i \gg \ell_W$ in the absence of instrument noise
takes the form
\begin{eqnarray}
\label{eqn:ttcov}
{\rm Cov}_{\el_i \el_j} & \equiv & 
\langle \hat C_{\ell_i} \hat C_{\ell_j} \rangle - 
\langle \hat C_{\ell_i} \rangle 
\langle \hat C_{\ell_j} \rangle  \\
&=&
 {2 C_{\ell_i}^2  \over (2\el_i+1)\Delta\ell_i \fsky}   \delta_{i j}
%\nonumber\\
%&&
+ \frac{\partial  \el_i^2 C_{\el_i} }{\partial \ln  \el_i}   \frac{\partial {\el_j^2} C_{\el_j} }{\partial \ln \el_j}\frac{ \sigma_\kappa^2}{\el_i^2 \el_j^2}  \nonumber\\
 &&+ \sum_{L}\Bigg[  {\partial C_{\el_i}\over \partial  C_{L}^{\phi\phi} } {\rm Cov}_{LL}^{\phi\phi,\phi\phi} 
{\partial C_{\el_j} \over \partial  C_{L}^{\phi\phi} } \Bigg], \nonumber
\end{eqnarray}
where $C_L^{\phi\phi}$ is the power spectrum of the lensing potential $\phi$ and $\sigma_\kappa$ is the rms fluctuation in the associated convergence field $\kappa = -\nabla^2 \phi/2$
 in the finite
survey area
\begin{equation}
\sigma_\kappa^2 =\frac{1}{A^2} \sum_{LM} |W_{LM}|^2 \frac{L^2(L+1)^2}{4} C_L^{\phi\phi}.
\label{eqn:sigmakappaallsky}
\end{equation}
Here $W_{LM}$ is the harmonic transform of the window.  
Finally the sample variance of $\hat C_L^{\phi\phi}$ in the survey area is given in the
Gaussian approximation as 
\begin{equation}
 {\rm Cov}_{LL}^{\phi\phi,\phi\phi} =  {2 \over (2L +1)\fsky}  [C_{L}^{\phi\phi}]^2 .
 \label{eqn:Covphiphi}
\end{equation}

The first term in Eq.~(\ref{eqn:ttcov}) is the usual connected or Gaussian contribution.     As we shall see explicitly in the
trispectrum derivation of \S \ref{sec:trispect}, the $\fsky$ or survey area factor $A=4\pi \fsky$ 
represents the fact that within the band $\Delta \ell_i$ only modes separated by more than $\ell_W$
are independent due to convolution by the window.
 
The second term is linear in $C_L^{\phi\phi}$ and represents the effect of the mean
convergence in the field.   Since  the mean
convergence dilates the whole field,  the power per logarithmic
interval  simply shifts in scale.  In Eq.~(\ref{eqn:ttcov}) we represent this power in the flat sky  limit as $\ell^2 C_\ell/2\pi$.
 While this approximation
requires corrections for the curved sky if $\ell \lesssim 60$,  
the covariance induced by the shift is much smaller than the Gaussian variance
term here  and the  error in the approximation has negligible impact.
 For $\sigma_\kappa \ll 1$, the
covariance then takes the form of the second term.  
Following Ref.~\cite{Takada:2013wfa}, we call this effect a super-sample covariance (SSC)
since the modes involved in producing the average convergence are on scales
larger than the survey.

Since this rms convergence $\sigma_\kappa$ rapidly declines with survey area, the SSC effect was omitted in the all-sky analysis of Ref.~\cite{BenoitLevy:2012va}.   Instead they introduced the third term  in Eq.~(\ref{eqn:ttcov})  which 
is higher order in $C_L^{\phi\phi}$ to describe the covariance
produced by   sampling fluctuations in the lensing potential power rather than individual modes.
Since this response is very smooth in $L$, we have omitted binning in $L$ for notational clarity. 
This Gaussian sample variance of Eq.~(\ref{eqn:Covphiphi}), like the first term, also scales with $\fsky^{-1}$.

\begin{figure}[tb]
\centering
    \includegraphics[width=3.4in]{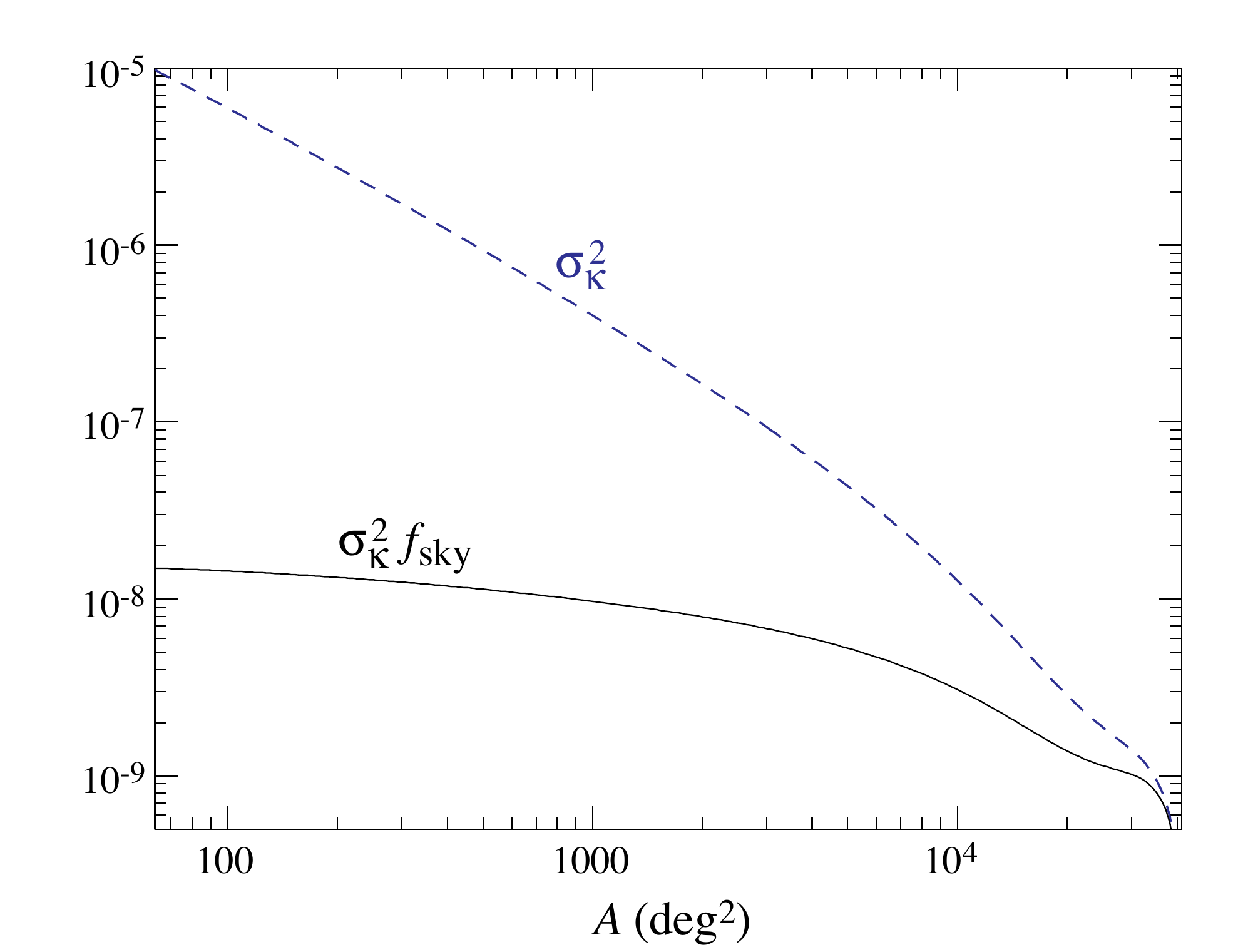}
    \caption{\footnotesize Variance in the mean convergence  $\sigma_\kappa^2$
    in a circular patch of sky of area $A$.    $\sigma_\kappa^{2}$ itself  falls nearly as white noise (blue dashed line) leaving the quantity relevant for comparing terms in the power spectrum covariance $\sigma_\kappa^{2}f_{\rm{sky}}$ (black, solid line) weakly dependent on survey area.}
    \label{fig:sigmafsky}
\end{figure}

Thus the relative contribution of the SSC compared with other
terms scales as $\sigma_\kappa^2 \fsky$ and its dependence on $\fsky$ or $A$ is
the relevant quantity to compute.   For example if
we consider a circular cap of polar angle $\theta_W$ then 
\begin{eqnarray}
W_{LM} & =& \sqrt{\frac{\pi}{2L+1}} [{P_{L-1}(x) -
P_{L+1}(x)}]\delta_{M0} ,
\end{eqnarray}
where $P_L$ is the Legendre polynomial with $x=\cos\theta_W$.   Note that 
\begin{equation}
\frac{4\pi}{A^2(2L+1)}\sum_M |W_{LM}|^2 = 
\begin{cases}
1,
& L \ll \ell_W \\
0, & L \gg \ell_W
\end{cases},
\end{equation}
where $A = 2\pi(1-x)$. 
As expected, the window function imposes a low pass filter on the total convergence power 
near the scale $\ell_W$.
Since $C_L^{\phi\phi}$ is approximately proportional to $L^{-4}$, the convergence power
is nearly white and  $\sigma_\kappa^2 
\propto \ell_W^2 \propto \fsky^{-1}$.   Thus 
\begin{equation}
\sigma_\kappa^2 f_{\rm sky} \approx \text{const.}
\end{equation} 
 and the importance of the SSC term is very weakly dependent
on the survey size.   We quantify these considerations in Fig.~\ref{fig:sigmafsky}
for a 6 parameter flat $\Lambda$CDM model that best fits the  Planck+WP+lensing data combination \footnote{\href{http://www.sciops.esa.int/wikiSI/planckpla}{Planck Explanatory Supplement}}:
\begin{eqnarray}
&& \{ \Omega_ch^2, \Omega_bh^2, 100 \theta_*, \tau, n_s, 10^9 A_s \} \nonumber\\
&& \hspace{0.3cm} = \{ 0.118, 0.0223, 1.04167,  0.0947, 0.968, 2.215\} ,
\label{eqn:parameters}
\end{eqnarray}
where $\theta_*$ is the angular size of the sound horizon %as defined by CAMB 
which
implies $h= 0.682$ and we also follow the Planck analysis in fixing $\Omega_\nu h^2=6.45\times 10^{-4}$ for consistency with
neutrino oscillation experiments.
Note that we include here only contributions for $L \ge 2$, which causes the sharp decline in $\sigma_\kappa^2 \fsky$ as $\fsky \rightarrow 1$.

\subsection{Trispectrum}
\label{sec:trispect}

The SSC term in Eq.~(\ref{eqn:ttcov}) can be more rigorously derived from the CMB trispectrum  \cite{Hu:2001fa} following
Ref.~\cite{Takada:2013wfa}.
  For simplicity, we employ the flat sky approximation in this section so that multipoles represent wavenumbers of a Fourier
transform.  
As discussed in the previous section, this should be a good approximation for band
powers where the covariance is important even if the survey itself is large enough to 
require a curved sky approach.   
In Fourier space, the window convolves modes in the spectrum
\begin{equation}
T_W(\bl) = \int \frac{d^2 \bl'}{(2\pi)^2} W(\bl-\bl') T(\bl').
\end{equation}
We can define an estimator of the temperature fluctuation field as
\begin{equation}
\hat C_{\el_i} = \frac{1}{A} \int_{\el \in \el_i} \frac{d^2 \el}{A_\el} T_W(\bl) T_W(-\bl),
\end{equation}
where  the integral is over modes in some band around $\el_i$ and $A_\el= 2\pi \el \Delta \el$.
Its expectation value
\begin{equation}
\langle \hat C_{\el_i}\rangle = \frac{1}{A} \int_{\el \in \el_i} \frac{d^2 \el}{A_\el}  \int \frac{d^2 \ell'}{(2\pi)^2} 
| W(\bl')|^2 C_{|\bl-\bl'|}
\end{equation}
 is unbiased %for $\ell_i \gg \ell_W$, where $\ell_W \sim 1/\sqrt{A}$ is the fundamental
%mode defined by the survey area,  
so long as $C_\ell$
has no features on  scales smaller than $\Delta \el_i$ since 
\begin{equation}\label{eqn:normalization}
\int \frac{d^2 \ell'}{(2\pi)^2} 
| W(\bl ')|^2  = A.
\end{equation}

Using the fact that the power spectrum and trispectrum, or connected 4-pt function,
are defined as
\begin{eqnarray}
\label{eqn:tri}
\langle T(\bl_1) T(\bl_2) \rangle &=& (2\pi)^2 \delta(\bl_{12}) C_{\el_1}, \\
\langle T(\bl_1) T(\bl_2) T(\bl_3) T(\bl_4) \rangle_c
&=& (2\pi)^2 \delta(\bl_{1234}) T(\bl_1,\bl_2,\bl_3,\bl_4) . \nonumber
\end{eqnarray}
where here and throughout ${\bl}_{1\ldots n} = {\bl}_1 + \ldots {\bl}_n$,
we can divide the resulting covariance of the estimators into two pieces
\begin{eqnarray}
{\rm Cov}_{\el_i \el_j}  &=& {\rm G}_{\el_i \el_j} + {\rm NG}_{\el_i \el_j}.
\end{eqnarray}
The first term is the disconnected or Gaussian contribution of Eq.~(\ref{eqn:ttcov})\begin{equation}
{\rm G}_{\el_i \el_j}
= \frac{ (2\pi)^2}{A A_\el} 2 C_{\ell_i}^2 \delta_{ij},
\end{equation}
%
%\am{should we underline that this can be understood from gaussian statistics as the inverse of the number of independent modes multiplying $C_{\ell_i}^2$ with $N_{modes} =\frac{A_{\ell}A}{(2\pi)^{2}}$ where $\frac{(2\pi)^{2}}{A}$ is the ``fundamental area'' in $\ell$ space? It is interesting but already mentioned in the second section}
%
as can be seen from taking $A=4 \pi \fsky$.   This also defines the number of effectively
independent $\ell$-modes 
\begin{equation}
N_\text{modes} =\frac{A_{\ell}A}{(2\pi)^{2}} = \frac{A_\ell}{\ell_W^2}
\end{equation}
corresponding to the $\Delta \ell < \ell_W$ criteria of the previous section.

The second term comes from the trispectrum
\begin{eqnarray}
{\rm NG}_{\el_i \el_j} &=& \frac{1}{A^2}  \int_{\el \in \el_i} \frac{d^2 \el}{A_\el}
 \int_{\el' \in \el_j} \frac{d^2 \el'}{A_{\el'}} \\
 && \int \left[ \prod_{a=1}^4 \frac{d^2 L_a}{(2\pi)^2}
 W(\bL_a) \right] (2\pi)^2\delta( \bL_{1234}) \nonumber\\
 &&T(\bl + \bL_1, -\bl + \bL_2, \bl' + \bL_3, -\bl + \bL_4). \nonumber
\end{eqnarray}
In the  limit that $\el_i \gg L_i$, we can relabel $\bl + \bL_1 \rightarrow \bl$, $\bl' + \bL_3 \rightarrow \bl'$, and the delta function condition sets $\bL=\bl_{12} = -\bl_{34}$. Since 
$W^2(\btheta)=W(\btheta)$, the convolution theorem
\begin{equation}
\int \frac{d^2 L_1}{(2\pi)^2} W(\bL_1) W(\bL-\bL_1) = W(\bL),
\end{equation}
can be used to express
\begin{eqnarray}
{\rm NG}_{\el_i \el_j}   &=& \frac{1}{A^2}  \int_{\el \in \el_i} \frac{d^2 \el}{A_\el}
 \int_{\el' \in \el_j} \frac{d^2 \el'}{A_{\el'} }
\int \frac{d^2 L}{(2\pi)^2} | W(\bL)|^2 \nonumber
\\
&& T(\bl,-\bl+\bL,\bl',-\bl'-\bL) .
\label{eqn:CovNG}
\end{eqnarray}
Note
that the window connects multipoles that are separated by less than its fundamental mode $\el_W$ and so the covariance no longer depends only on degenerate quadrilaterals through
$T(\bl,-\bl,\bl',-\bl')$ but rather squeezed quadrilaterals of Eq.~(\ref{eqn:CovNG}).

To linear order in $C_\ell^{\phi\phi}$, the lensing trispectrum is given by \cite{Hu:2001fa}
\begin{equation}
T(\bl_1,\bl_2,\bl_3,\bl_4) = C_{\el_1} C_{\el_3} C^{\phi\phi}_{\el_{12}} 
\left( \bl_{12} \cdot \bl_1 \right) 
\left(\bl_{34}\cdot \bl_3 \right)+ {\rm perm.}
\end{equation}
where ``perm." means all permutations of the $\bl_i$.    
In the relevant squeezed quadrilateral limit
$L\ll \el_1$, we can  expand
\begin{eqnarray}
\el_2 \approx  \el_1 - \frac{\bl_1\cdot \bl_{12} }{\el_1}
% ,\quad 
%\el_4 \approx  \el_3 - \frac{\bl_3 \cdot \bl_{34}} {\el_3} ,
\end{eqnarray}
so that
\begin{eqnarray}
C_{\el_2} &\approx & C_{\el_1}  - \frac{\partial C_\el }{\partial \ln \el} \bigg|_{\el_1} \frac{\bl_1\cdot \bl_{12} }{\el_1^2}
%C_{\el_4} &\approx & C_{\el_3} -  \frac{\partial  C_\el }{\partial \ln \el} \bigg|_{\el_3}  \frac{\bl_3 \cdot \bl_{34}}{\el_3^2 } ,
\end{eqnarray} 
and rewrite
\begin{eqnarray}
\bl_{12} \cdot \bl_2& =& -\bl_{12} \cdot \bl_1+  \el_{12}^2, 
%\bl_{34} \cdot \bl_4 &= &-\bl_{34} \cdot \bl_3+  \el_{34}^2.
\end{eqnarray}
and similarly for $1 \rightarrow 3$ and $2 \rightarrow 4$.
With these replacements in Eq.~(\ref{eqn:CovNG}), we obtain 
\begin{equation}
{\rm NG}_{\el_i \el_j} = \frac{\partial {\el_i^2} C_{\el_i} }{\partial \ln \el_i} \frac{\partial  \el_j^2 C_{\el_j} }{\partial \ln \el_j} \frac{\sigma_\kappa^2}{\el_i^2\el_j^2} ,
\end{equation}
where
\begin{equation}\label{eqn:sigma}
\sigma_\kappa^2 =\frac{1}{A^2} \int \frac{d^2 L}{(2\pi)^2} \frac{L^4}{4} C_L^{\phi\phi} |W(\bL)|^2,
\end{equation}
which is the flat-sky limit of Eq.~(\ref{eqn:sigmakappaallsky}). Thus the trispectrum to linear order in $C_L^{\phi\phi}$ yields the SSC term of Eq.~(\ref{eqn:ttcov}).  Because this term vanishes for all-sky measurements where $L \rightarrow 0$ , or more properly becomes
indistinguishable from a change in the background cosmology, Ref.~\cite{BenoitLevy:2012va}
found that the dominant term is higher order in $C_L^{\phi\phi}$.    Here we consider the
SSC effect for smaller survey fields.

\subsection{Polarization}
\label{sec:pol}

The full covariance matrix of temperature and  $E$, $B$ polarization power spectra takes
a similar form
\begin{eqnarray}
{\rm Cov}^{W X, Y Z}_{\ell_i \ell_j}  &\equiv&
\langle \hat C_{\ell_i}^{WX} \hat C_{\ell_j}^{YZ} \rangle - 
\langle \hat C_{\ell_i}^{WX} \rangle 
\langle \hat C_{\ell_j}^{YZ} \rangle \nonumber\\
&=& {\rm G}^{W X, Y Z}_{\ell_i \ell_j} + {\rm NG}^{W X, Y Z}_{\ell_i \ell_j} ,
\end{eqnarray}
where $\{ W, X, Y, Z \} \in \{ T, E, B \}$. 
The Gaussian contribution is
\begin{equation}
{\rm G}^{W X, Y Z}_{\ell_i \ell_j}  = 
 { C_{\ell_i}^{WY}C_{\ell_i}^{XZ} + C_{\ell_i}^{WZ} C_{\ell_i}^{XY} \over (2\ell_i+1)\Delta\ell_i \fsky} \delta_{i j} ,
\label{eqn:Gcov}
\end{equation}
where $C_\el^{TT}= C_\el$,
and the SSC or linear in $C_L^{\phi\phi}$ term is
\begin{equation}
{\rm NG}^{W X, Y Z}_{\ell_i \ell_j} = 
 \frac{\partial  {\el_i^2} C_{\el_i}^{WX}  }{\partial \ln \el_i} \frac{\partial  \el_j^2 C_{\el_j}^{YZ} }{\partial \ln \el_j}  \frac{\sigma_\kappa^2}{\el_i^2 \el_j^2}  + {\cal O}(C_L^{\phi\phi})^2.
 \label{eqn:SSCpol}
\end{equation}
For the cases involving $TT$, $EE$, and $TE$, this expression can be directly derived
from the leading order trispectrum \cite{Hu:2000ee,Okamoto:2002ik} in the same way as in the previous section.
For $BB$ from lensing, the contribution is higher than leading order since $C_\ell^{BB}$  is itself ${\cal O}(C_L^{\phi\phi})$.   Its form can be inferred from the fact that the
average convergence in the field lenses CMB polarization that is itself lensed into $B$
modes by
small scale perturbations.   This logic parallels that of the refinement
\cite{Hanson:2010rp,Lewis:2011fk} of lensing reconstruction
estimators \cite{Hu:2001tn,Okamoto03} where terms that are higher order in lensing are accounted for by replacing
the unlensed CMB power spectra with the lensed CMB power spectra.

For completeness, we include the ${\cal O}(C_L^{\phi\phi})^2$ terms introduced in 
Ref.~\cite{BenoitLevy:2012va}.
For the $BB$ power spectrum,
\begin{eqnarray}
{\rm NG}^{BB,BB}_{\ell_i \ell_j } & =& 
\ldots  + \sum_{L}
{\partial C_{\ell_i }^{BB}  \over \partial  C_{L}^{\tilde E\tilde E} } {\rm Cov}_{LL}^{\tilde E\tilde E,\tilde E\tilde E}
{\partial C_{\ell_j}^{BB} \over \partial  C_{L}^{\tilde E\tilde E} }  \nonumber\\ &&
 + \sum_{L}  {\partial C_{\ell_i}^{BB}\over \partial  C_{L}^{\phi\phi} } {\rm Cov}_{LL}^{\phi\phi,\phi\phi} {\partial C_{\ell_j}^{BB} \over \partial  C_{L}^{\phi\phi} }  ,
\label{eqn:BBBB}
\end{eqnarray}
where tildes denote the unlensed power spectrum and its covariance  follows the Gaussian prescription of Eq.~(\ref{eqn:Gcov}).
Here and below ``$\ldots$'' denotes the linear SSC term of Eq.~(\ref{eqn:SSCpol}).
For $WX=BB$ and $\{ Y, Z \} \in \{T , E\}$,
\begin{eqnarray}
{\rm NG}^{BB,YZ}_{\ell_i \ell_j }& =& \ldots +
 \sum_{L}
{\partial C_{\ell_i}^{BB}  \over \partial  C_{L}^{\tilde E\tilde E} } 
{\rm Cov}^{\tilde E\tilde E,\tilde Y\tilde Z}_{LL}
{\partial C_{\ell_j}^{YZ} \over \partial  C_{L}^{\tilde Y \tilde Z} } 
 \nonumber\\ &&
 + \sum_{L}   {\partial C_{\ell_i}^{BB}\over \partial  C_{L}^{\phi\phi} } {\rm Cov}_{LL}^{\phi\phi,\phi\phi}  {\partial C_{\ell_j}^{YZ} \over \partial  C_{L}^{\phi\phi} }  ,
 \label{eqn:BBXY}
\end{eqnarray}
and for $\{ W, X \} \in \{ T, E\}$  as well
\begin{equation}
{\rm NG}^{WX,YZ}_{\ell_i \ell_j}  = 
\ldots + \sum_{L} {\partial C_{\ell_i}^{WX}\over \partial  C_{L}^{\phi\phi} } {\rm Cov}_{LL}^{\phi\phi,\phi\phi} 
{\partial C_{\ell_j}^{YZ} \over \partial  C_{L}^{\phi\phi} }.\,\,\,\,\,
\label{eqn:XYWZ}
\end{equation}
The impact of these other terms on parameter estimation was addressed in 
Ref.~\cite{BenoitLevy:2012va} and so we focus on that of the SSC term next.

\section{Signal vs. Noise}
\label{sec:impact}

A simple way 
 to assess the impact of the SSC effect is to consider it as part of 
the signal rather than the noise.   As noise, SSC introduces a covariance because the whole power spectrum dilates with the average convergence in the field $\bar \kappa$ whose variance $\sigma_\kappa^2$ makes band powers covary field-to-field.   As signal, $\bar \kappa$ is simply added to the cosmological parameters ${\bf p}$ in the likelihood analysis of
an individual field
\begin{equation}
\hat C_\el^{XY}({\bf p};\kappa)  =  C_\el^{ XY}({\bf p})-  \frac{\partial  \el^2  C_l^{XY}(\bf p)}{\partial \ln \el} \frac{\bar \kappa }{\ell^2},
\end{equation}
which is subject to a Gaussian prior given $\sigma_\kappa^2$.   

To determine whether the SSC effect is important to include for any cosmological parameter
set, it is sufficient to test its impact on an artificial  fully-degenerate parameter ${\bf p}=s$ that also
dilates scales in the power spectra
\begin{equation}
C_\el^{XY}(s)  = \bar C_\el^{\rm XY} + \frac{\partial  \el^2 \bar C_l^{XY} }{\partial \ln \el} \frac{s}{\ell^2} .
\end{equation}
If for a given experiment
$\sigma_s < \sigma_\kappa$ then  $\bar \kappa$ should in principle be marginalized in the analysis or SSC included in the covariance.  
Here $\bar C_\ell^{\rm XY}$ represents some fixed fiducial power spectrum which we take 
to be the Planck best fit model described in \S \ref{sec:temp}.  It is interesting to note that
if the inflationary power spectrum contains features of width $\Delta \ell$ which are finer than the acoustic
spacing $\ell_W < \Delta \ell \lesssim 300$, then the impact of SSC can be greatly enhanced
\cite{Miranda:2013wxa}.

We can estimate $\sigma_s$  with the Fisher technique.   For an arbitrary set of 
parameters ${\bf p}$, $\sigma^2_{p_\mu}= ({\bf F}^{-1})_{\mu\mu}$ where the Fisher matrix
\begin{equation}
F_{\mu\nu} = \sum_{ij} \sum_{WX, YZ} \frac{\partial C_{\el_i}^{WX}}{\partial p_\mu} \left( {\rm G}_{\el_i \el_j}^{WX,YZ}\right)^{-1}
\frac{\partial C_{\el_j}^{YZ}}{\partial p_\nu} .
\label{eqn:fisher}
\end{equation}
   Here we specialize to a single
parameter $s$ and  the  Gaussian covariance of Eq.~(\ref{eqn:Gcov}).     This is the Gaussian sample variance limit.   

\begin{figure}[tb]
\centering
    \includegraphics[width=3.6in]{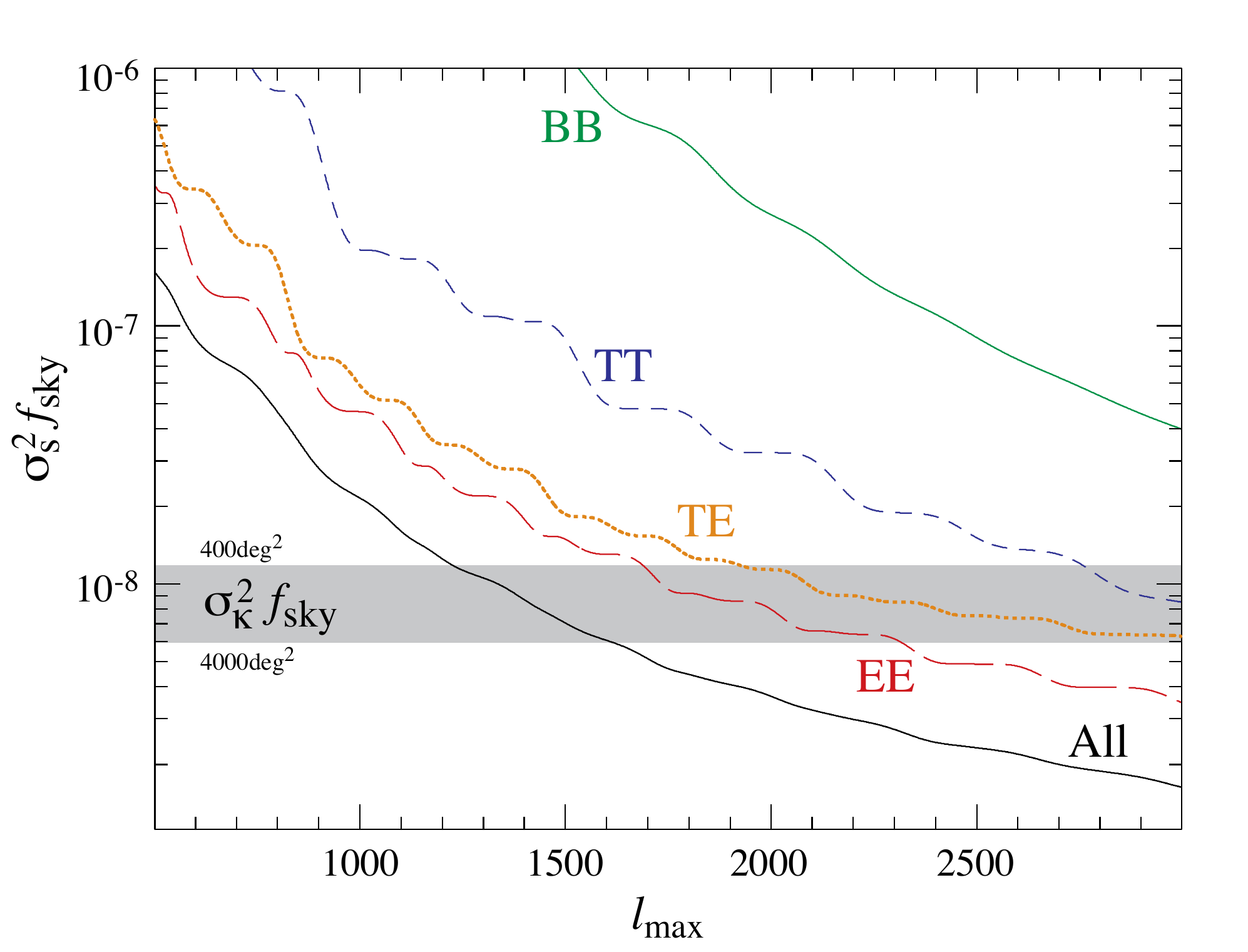}
    \caption{\footnotesize  Gaussian sample variance limited constraints on a parameter $s$  that
    dilates the CMB power spectra similar to $\ln \theta_*$ in the acoustic regime.
    When the expected variance $\sigma_s^2$ is comparable to or smaller  that of the mean convergence 
    in the field $\sigma_\kappa^2$, the SSC effect should be included in the analysis.   For a wide range of sky coverage $400 \le A( {\rm deg}^2) \le 4000$, $\sigma_\kappa^2\fsky$ is within the shaded band indicating that
    SSC is important for such measurements if they are sample variance limited to at least $\ell_{\rm max} \sim 1200-1600$.}
    \label{fig:error}
\end{figure}

In Fig.~\ref{fig:error}, we show the quantity $\sigma_s^2 \fsky$ in the limit of bins that
are much finer than the acoustic scale $\Delta \ell \ll 300$ summed out to $\ell_i \le \ell_{\rm max}$.   This should be compared with $\sigma_\kappa^2 \fsky$ which we also 
show for comparison for the range of $400 \le A (\text{deg}^2) \le 4000$ and a circular patch from Fig.~\ref{fig:sigmafsky}.  Note that
for  sample variance limited measurements of all power spectra, the two cross
between $1200 \lesssim \ell_{\rm max} \lesssim 1600$.   Thus the SSC effect should
be modeled for surveys which are sample variance limited to at least these multipoles.

We also show in Fig.~\ref{fig:error} the individual
contributions for each type of spectrum.   The measurement of $s$ mainly 
reflects the $EE$ and $TE$ power spectra given the more prominent acoustic features in 
these spectra.   For the $TT$ spectrum alone, this crossing point is pushed to
$\ell_{\rm max} \gtrsim 2800$ and a real survey is likely to be foreground limited
before this point.  For the $BB$ spectrum there is negligible impact even for sample variance limited measurements given its smooth form.

To further make these considerations concrete we can include instrument noise $N_\ell^{XY}$
 by replacing in the Gaussian covariance
of  Eq.~(\ref{eqn:Gcov})
\begin{equation}
C_\ell^{XY} \rightarrow C_\ell^{XY} + N_\ell^{XY},
\end{equation}
where
\begin{equation}
N_\ell^{TT}= \frac{1}{2}N_\ell^{EE}  = \frac{1}{2}N_\ell^{BB} = \Delta_T^2 e^{\ell(\ell+1)\theta^{2}_\text{FWHM}/8\ln 2},
\end{equation}
with all other $N_\ell^{XY}=0$. Here $\Delta_T$ is the noise level in $\mu$K-radians and
$\theta_\text{FWHM}$ is the full-width half-max of the beam in radians.
We also restrict the bands to $\ell_{\rm max}\le 2000$, $3000$ beyond which the primary
anisotropy may be too difficult to extract from foregrounds and secondaries. 
In Tab.~\ref{tab:noises}, we show that for a typical second generation (2G) and
third generation (3G) survey,
the SSC variance $\sigma_\kappa^2$ exceeds $\sigma_s^2$ by a factor of 2-3 for $\ell_{\rm max}=2000-3000$.  
This indicates that the SSC effect should be included in the analysis of these surveys. 

While $\sigma_\kappa^2/\sigma_s^2$ determines whether SSC is important at all, its
impact on physical cosmological parameters depends on whether or not the
given parameter is degenerate  with a pure shift $s$. 
We can use the Fisher technique to study the SSC impact on parameter estimation by appending $\bar\kappa$ to these
parameters and adding a prior 
\begin{equation}
F_{\mu\nu}^{\rm prior}=\sigma_{\kappa}^{-2} \delta_{\bar\kappa,\mu} \delta_{\bar\kappa,\nu}.
\end{equation}
The impact of SSC can be determined by comparing the change in parameter errors upon marginalizing
$\bar \kappa$ vs fixing $\bar \kappa=0$.

   In the $\Lambda$CDM parameter space  defined by the parameters in Eq.~(\ref{eqn:parameters}), only the angular size of the sound horizon $\theta_*$ is nearly degenerate with $s$ and hence only its variance degrades significantly.
In Tab.~\ref{tab:noises} we quantify this degradation as
\begin{equation}
R_{\theta_*} \equiv \frac{\sigma_{\ln\theta_*}^2 }{\sigma_{\ln\theta_*}^2\big|_{\bar \kappa=0}}
\le  1+\frac{\sigma_\kappa^2}{\sigma^2_{s}}.
%\approx \frac{\sigma_{\theta_*}^2 }{\sigma_{\theta_*}^2\big|_{\bar \kappa=0}}.
\end{equation}
The inequality follows from the fact that at low $\ell$ 
the power spectrum is not dominated by acoustic features and more importantly, at high $\ell$ lensing effects
enter so that $\sigma_{\ln\theta_*}^2\ge
\sigma_s^2$.
Even so, the degradation in the variance of $\theta_*$ is $R_{\theta_*} \sim 2-3$   for the 
2G and 3G experiments.  
{We have also conducted this test using the covariance approach by adding only the
SSC term of Eq.~(\ref{eqn:SSCpol}) to the Gaussian error and omitting $\bar\kappa$ as
a parameter.  This method obtains nearly identical results for the degradation in $\sigma_{\ln \theta_*}^2$.}

For estimation purposes, it is useful to note that
given the degeneracy between $\ln \theta_*$ and $\kappa$ this degradation factor is to good
approximation 
\begin{eqnarray}
R_{\theta_*} \approx 1+\frac{\sigma_\kappa^2}{\sigma_{\ln\theta_*}^2\big|_{\bar \kappa=0}} .
\end{eqnarray}
For example, this expression approximates values in Tab.~\ref{tab:noises} at the 1\% level.

We can use this approximation to estimate the SSC effect for the Planck data.   Taking the smallest patch of
sky $\fsky=0.37$ from which the power spectrum is measured to maximize the SSC effect and
crudely assuming a circular patch, we obtain $\sigma_\kappa \sim 7 \times 10^{-5}$.   
This should be
compared with the quoted precision on the acoustic scale $\sigma_{\ln \theta_*}=5.8 \times 10^{-4}$.   
We thus conclude that the SSC effect is unimportant for the analysis of the Planck data.

On the other hand, its omission in the 2G and 3G cases may lead to seemingly discrepant
results for $\theta_*$ when compared with either Planck results  or with each other.  Discrepant results on $\theta_*$ have little
impact on CMB measurements of fundamental parameters through the distance to recombination
due to uncertainties in the physical scale of the sound horizon.  For example, the errors on the Hubble constant in a flat $\Lambda$CDM model are dominated by errors on
$\Omega_{c}h^2$ not $\theta_*$.  However they may lead to erroneous results when compared to measurements of the same scale through baryon acoustic oscillations.

\begin{table}
\centering
  \begin{tabular}{ |c | c | c|c| c| c| c| c|}
    \hline
    &$\Delta_{T}$  & $\theta_\text{FWHM}$ & $A$ & \multicolumn{2}{|c|}{1+$\sigma_\kappa^2/\sigma_s^2(\ell_{\rm max})$} &\multicolumn{2}{|c| }{$R_{\theta_*}(\ell_{\rm max})$}\\ 
    & ($\mu\text{K}'$) &  (arcmin) & (deg$^2$) & 2000 & 3000 & 2000 & 3000 \\\hline
    2G & 12 & $1'$& 500 & 2.7 &3.5& 2.4  & 2.6 \\ \hline
    3G & 3.5 & $1'$ & 2500 & 2.8 & 4.1&2.5 & 3.0 \\
    \hline
  \end{tabular}
  \caption{\footnotesize Relative importance of SSC for example experimental specifications.  $1+\sigma^2_\kappa/\sigma_s^2$ gives the maximal degradation in variance
  for a parameter that is degenerate with the shift due to lensing whereas $R_{\theta_*}$ gives
  the same for the angular size of the sound horizon. }
\label{tab:noises}	
\end{table}

\section{Discussion}
\label{sec:discussion}

We have explored the effect of CMB lensing by super sample modes on power spectra
measurements within a finite sky sample.   The resulting  field-to-field fluctuations in the mean 
convergence dilates the intrinsic scales in the survey and coherently shifts the measured
power spectra with respect to the ensemble or all-sky mean.    The effect can be 
simply encapsulated  as an extra fully covarying SSC contribution to the
covariance matrix, as described by the lensing trispectrum, or as an additional parameter,
the mean convergence, that must be included in parameter estimation.

By comparing the ability of a given experiment to measure an overall change in 
angular scale to the expected rms mean convergence, we provide a simple test for 
when the SSC effect should be included in data analysis.   This criterion is met  for surveys that precisely measure
acoustic polarization features deep into the damping tail at multipoles of $\ell \gtrsim 1500$ with less than $\sim 10\%$ of sky.   If this effect is omitted in the analysis then different
surveys may measure seemingly conflicting values for the angular scale of the sound
horizon $\theta_*$ any one of which may then provide erroneous fundamental cosmological parameters when compared
to external measurements of the same standard ruler from baryon acoustic oscillations.

\acknowledgements We thank Clarence Chang for providing the SPTPol and SPT-3G  target specifications for our 2G and 3G test cases and Kendrick Smith for useful conversations. 
WH was supported by 
 by U.S.~Dept.\ of Energy
 contract DE-FG02-13ER41958,
  the
 Kavli Institute for Cosmological Physics at the University of
 Chicago through grants NSF PHY-0114422 and NSF PHY-0551142 and the David and Lucile Packard Foundation. ABL Acknowledges support from the Leverhulme Trust and the Science and Technology Facilities Council.

\bibliography{SSLens}

\begin{thebibliography}{21}%
\makeatletter
\providecommand \@ifxundefined [1]{%
 \@ifx{#1\undefined}
}%
\providecommand \@ifnum [1]{%
 \ifnum #1\expandafter \@firstoftwo
 \else \expandafter \@secondoftwo
 \fi
}%
\providecommand \@ifx [1]{%
 \ifx #1\expandafter \@firstoftwo
 \else \expandafter \@secondoftwo
 \fi
}%
\providecommand \natexlab [1]{#1}%
\providecommand \enquote  [1]{``#1''}%
\providecommand \bibnamefont  [1]{#1}%
\providecommand \bibfnamefont [1]{#1}%
\providecommand \citenamefont [1]{#1}%
\providecommand \href@noop [0]{\@secondoftwo}%
\providecommand \href [0]{\begingroup \@sanitize@url \@href}%
\providecommand \@href[1]{\@@startlink{#1}\@@href}%
\providecommand \@@href[1]{\endgroup#1\@@endlink}%
\providecommand \@sanitize@url [0]{\catcode `\\12\catcode `\$12\catcode
  `\&12\catcode `\#12\catcode `\^12\catcode `\_12\catcode `\%12\relax}%
\providecommand \@@startlink[1]{}%
\providecommand \@@endlink[0]{}%
\providecommand \url  [0]{\begingroup\@sanitize@url \@url }%
\providecommand \@url [1]{\endgroup\@href {#1}{\urlprefix }}%
\providecommand \urlprefix  [0]{URL }%
\providecommand \Eprint [0]{\href }%
\providecommand \doibase [0]{http://dx.doi.org/}%
\providecommand \selectlanguage [0]{\@gobble}%
\providecommand \bibinfo  [0]{\@secondoftwo}%
\providecommand \bibfield  [0]{\@secondoftwo}%
\providecommand \translation [1]{[#1]}%
\providecommand \BibitemOpen [0]{}%
\providecommand \bibitemStop [0]{}%
\providecommand \bibitemNoStop [0]{.\EOS\space}%
\providecommand \EOS [0]{\spacefactor3000\relax}%
\providecommand \BibitemShut  [1]{\csname bibitem#1\endcsname}%
\let\auto@bib@innerbib\@empty
%</preamble>
\bibitem [{\citenamefont {Das}\ \emph {et~al.}(2011)\citenamefont {Das},
  \citenamefont {Sherwin}, \citenamefont {Aguirre}, \citenamefont {Appel},
  \citenamefont {Bond} \emph {et~al.}}]{Das:2011ak}%
  \BibitemOpen
  \bibfield  {author} {\bibinfo {author} {\bibfnamefont {S.}~\bibnamefont
  {Das}}, \bibinfo {author} {\bibfnamefont {B.~D.}\ \bibnamefont {Sherwin}},
  \bibinfo {author} {\bibfnamefont {P.}~\bibnamefont {Aguirre}}, \bibinfo
  {author} {\bibfnamefont {J.~W.}\ \bibnamefont {Appel}}, \bibinfo {author}
  {\bibfnamefont {J.~R.}\ \bibnamefont {Bond}},  \emph {et~al.},\ }\href
  {\doibase 10.1103/PhysRevLett.107.021301} {\bibfield  {journal} {\bibinfo
  {journal} {Phys.Rev.Lett.}\ }\textbf {\bibinfo {volume} {107}},\ \bibinfo
  {pages} {021301} (\bibinfo {year} {2011})},\ \Eprint
  {http://arxiv.org/abs/1103.2124} {arXiv:1103.2124 [astro-ph.CO]} \BibitemShut
  {NoStop}%
%%CITATION = ARXIV:1103.2124;%%
\bibitem [{\citenamefont {Das}\ \emph {et~al.}(2013)\citenamefont {Das},
  \citenamefont {Louis}, \citenamefont {Nolta}, \citenamefont {Addison},
  \citenamefont {Battistelli} \emph {et~al.}}]{Das:2013zf}%
  \BibitemOpen
  \bibfield  {author} {\bibinfo {author} {\bibfnamefont {S.}~\bibnamefont
  {Das}}, \bibinfo {author} {\bibfnamefont {T.}~\bibnamefont {Louis}}, \bibinfo
  {author} {\bibfnamefont {M.~R.}\ \bibnamefont {Nolta}}, \bibinfo {author}
  {\bibfnamefont {G.~E.}\ \bibnamefont {Addison}}, \bibinfo {author}
  {\bibfnamefont {E.~S.}\ \bibnamefont {Battistelli}},  \emph {et~al.},\
  }\href@noop {} {\  (\bibinfo {year} {2013})},\ \Eprint
  {http://arxiv.org/abs/1301.1037} {arXiv:1301.1037 [astro-ph.CO]} \BibitemShut
  {NoStop}%
%%CITATION = ARXIV:1301.1037;%%
\bibitem [{\citenamefont {Keisler}\ \emph {et~al.}(2011)\citenamefont
  {Keisler}, \citenamefont {Reichardt}, \citenamefont {Aird}, \citenamefont
  {Benson}, \citenamefont {Bleem} \emph {et~al.}}]{Keisler:2011aw}%
  \BibitemOpen
  \bibfield  {author} {\bibinfo {author} {\bibfnamefont {R.}~\bibnamefont
  {Keisler}}, \bibinfo {author} {\bibfnamefont {C.}~\bibnamefont {Reichardt}},
  \bibinfo {author} {\bibfnamefont {K.}~\bibnamefont {Aird}}, \bibinfo {author}
  {\bibfnamefont {B.}~\bibnamefont {Benson}}, \bibinfo {author} {\bibfnamefont
  {L.}~\bibnamefont {Bleem}},  \emph {et~al.},\ }\href {\doibase
  10.1088/0004-637X/743/1/28} {\bibfield  {journal} {\bibinfo  {journal}
  {Astrophys.J.}\ }\textbf {\bibinfo {volume} {743}},\ \bibinfo {pages} {28}
  (\bibinfo {year} {2011})},\ \Eprint {http://arxiv.org/abs/1105.3182}
  {arXiv:1105.3182 [astro-ph.CO]} \BibitemShut {NoStop}%
%%CITATION = ARXIV:1105.3182;%%
\bibitem [{\citenamefont {van Engelen}\ \emph {et~al.}(2012)\citenamefont {van
  Engelen}, \citenamefont {Keisler}, \citenamefont {Zahn}, \citenamefont
  {Aird}, \citenamefont {Benson} \emph {et~al.}}]{vanEngelen:2012va}%
  \BibitemOpen
  \bibfield  {author} {\bibinfo {author} {\bibfnamefont {A.}~\bibnamefont {van
  Engelen}}, \bibinfo {author} {\bibfnamefont {R.}~\bibnamefont {Keisler}},
  \bibinfo {author} {\bibfnamefont {O.}~\bibnamefont {Zahn}}, \bibinfo {author}
  {\bibfnamefont {K.}~\bibnamefont {Aird}}, \bibinfo {author} {\bibfnamefont
  {B.}~\bibnamefont {Benson}},  \emph {et~al.},\ }\href {\doibase
  10.1088/0004-637X/756/2/142} {\bibfield  {journal} {\bibinfo  {journal}
  {Astrophys.J.}\ }\textbf {\bibinfo {volume} {756}},\ \bibinfo {pages} {142}
  (\bibinfo {year} {2012})},\ \Eprint {http://arxiv.org/abs/1202.0546}
  {arXiv:1202.0546 [astro-ph.CO]} \BibitemShut {NoStop}%
%%CITATION = ARXIV:1202.0546;%%
\bibitem [{\citenamefont {Hanson}\ \emph {et~al.}(2013)\citenamefont {Hanson}
  \emph {et~al.}}]{Hanson:2013hsb}%
  \BibitemOpen
  \bibfield  {author} {\bibinfo {author} {\bibfnamefont {D.}~\bibnamefont
  {Hanson}} \emph {et~al.} (\bibinfo {collaboration} {SPTpol Collaboration}),\
  }\href {\doibase 10.1103/PhysRevLett.111.141301} {\bibfield  {journal}
  {\bibinfo  {journal} {Phys.Rev.Lett.}\ }\textbf {\bibinfo {volume} {111}},\
  \bibinfo {pages} {141301} (\bibinfo {year} {2013})},\ \Eprint
  {http://arxiv.org/abs/1307.5830} {arXiv:1307.5830 [astro-ph.CO]} \BibitemShut
  {NoStop}%
%%CITATION = ARXIV:1307.5830;%%
\bibitem [{\citenamefont {Ade}\ \emph {et~al.}(2013{\natexlab{a}})\citenamefont
  {Ade} \emph {et~al.}}]{Ade:2013tya}%
  \BibitemOpen
  \bibfield  {author} {\bibinfo {author} {\bibfnamefont {P.}~\bibnamefont
  {Ade}} \emph {et~al.} (\bibinfo {collaboration} {Polarbear Collaboration}),\
  }\href@noop {} {\  (\bibinfo {year} {2013}{\natexlab{a}})},\ \Eprint
  {http://arxiv.org/abs/1312.6646} {arXiv:1312.6646 [astro-ph.CO]} \BibitemShut
  {NoStop}%
%%CITATION = ARXIV:1312.6646;%%
\bibitem [{\citenamefont {Ade}\ \emph {et~al.}(2013{\natexlab{b}})\citenamefont
  {Ade} \emph {et~al.}}]{Ade:2013tyw}%
  \BibitemOpen
  \bibfield  {author} {\bibinfo {author} {\bibfnamefont {P.}~\bibnamefont
  {Ade}} \emph {et~al.} (\bibinfo {collaboration} {Planck Collaboration}),\
  }\href@noop {} {\  (\bibinfo {year} {2013}{\natexlab{b}})},\ \Eprint
  {http://arxiv.org/abs/1303.5077} {arXiv:1303.5077 [astro-ph.CO]} \BibitemShut
  {NoStop}%
%%CITATION = ARXIV:1303.5077;%%
\bibitem [{\citenamefont {{Smith}}\ \emph {et~al.}(2006)\citenamefont
  {{Smith}}, \citenamefont {{Hu}},\ and\ \citenamefont
  {{Kaplinghat}}}]{Smith06}%
  \BibitemOpen
  \bibfield  {author} {\bibinfo {author} {\bibfnamefont {K.~M.}\ \bibnamefont
  {{Smith}}}, \bibinfo {author} {\bibfnamefont {W.}~\bibnamefont {{Hu}}}, \
  and\ \bibinfo {author} {\bibfnamefont {M.}~\bibnamefont {{Kaplinghat}}},\
  }\href {\doibase 10.1103/PhysRevD.74.123002} {\bibfield  {journal} {\bibinfo
  {journal} {Phys. Rev. D}\ }\textbf {\bibinfo {volume} {74}},\ \bibinfo
  {pages} {123002} (\bibinfo {year} {2006})},\ \Eprint
  {http://arxiv.org/abs/arXiv:astro-ph/0607315} {arXiv:astro-ph/0607315}
  \BibitemShut {NoStop}%
\bibitem [{\citenamefont {Benoit-Levy}\ \emph {et~al.}(2012)\citenamefont
  {Benoit-Levy}, \citenamefont {Smith},\ and\ \citenamefont
  {Hu}}]{BenoitLevy:2012va}%
  \BibitemOpen
  \bibfield  {author} {\bibinfo {author} {\bibfnamefont {A.}~\bibnamefont
  {Benoit-Levy}}, \bibinfo {author} {\bibfnamefont {K.~M.}\ \bibnamefont
  {Smith}}, \ and\ \bibinfo {author} {\bibfnamefont {W.}~\bibnamefont {Hu}},\
  }\href {\doibase 10.1103/PhysRevD.86.123008} {\bibfield  {journal} {\bibinfo
  {journal} {Phys.Rev.}\ }\textbf {\bibinfo {volume} {D86}},\ \bibinfo {pages}
  {123008} (\bibinfo {year} {2012})},\ \Eprint {http://arxiv.org/abs/1205.0474}
  {arXiv:1205.0474 [astro-ph.CO]} \BibitemShut {NoStop}%
%%CITATION = ARXIV:1205.0474;%%
\bibitem [{\citenamefont {Takada}\ and\ \citenamefont
  {Hu}(2013)}]{Takada:2013wfa}%
  \BibitemOpen
  \bibfield  {author} {\bibinfo {author} {\bibfnamefont {M.}~\bibnamefont
  {Takada}}\ and\ \bibinfo {author} {\bibfnamefont {W.}~\bibnamefont {Hu}},\
  }\href {\doibase 10.1103/PhysRevD.87.123504} {\bibfield  {journal} {\bibinfo
  {journal} {Phys.Rev.}\ }\textbf {\bibinfo {volume} {D87}},\ \bibinfo {pages}
  {123504} (\bibinfo {year} {2013})},\ \Eprint {http://arxiv.org/abs/1302.6994}
  {arXiv:1302.6994 [astro-ph.CO]} \BibitemShut {NoStop}%
%%CITATION = ARXIV:1302.6994;%%
\bibitem [{\citenamefont {Hamilton}\ \emph {et~al.}(2006)\citenamefont
  {Hamilton}, \citenamefont {Rimes},\ and\ \citenamefont
  {Scoccimarro}}]{Hamilton:2005dx}%
  \BibitemOpen
  \bibfield  {author} {\bibinfo {author} {\bibfnamefont {A.~J.}\ \bibnamefont
  {Hamilton}}, \bibinfo {author} {\bibfnamefont {C.~D.}\ \bibnamefont {Rimes}},
  \ and\ \bibinfo {author} {\bibfnamefont {R.}~\bibnamefont {Scoccimarro}},\
  }\href {\doibase 10.1111/j.1365-2966.2006.10709.x} {\bibfield  {journal}
  {\bibinfo  {journal} {Mon.Not.Roy.Astron.Soc.}\ }\textbf {\bibinfo {volume}
  {371}},\ \bibinfo {pages} {1188} (\bibinfo {year} {2006})},\ \Eprint
  {http://arxiv.org/abs/astro-ph/0511416} {arXiv:astro-ph/0511416 [astro-ph]}
  \BibitemShut {NoStop}%
%%CITATION = ASTRO-PH/0511416;%%
\bibitem [{\citenamefont {Li}\ \emph {et~al.}(2014)\citenamefont {Li},
  \citenamefont {Hu},\ and\ \citenamefont {Takada}}]{Li:2014sga}%
  \BibitemOpen
  \bibfield  {author} {\bibinfo {author} {\bibfnamefont {Y.}~\bibnamefont
  {Li}}, \bibinfo {author} {\bibfnamefont {W.}~\bibnamefont {Hu}}, \ and\
  \bibinfo {author} {\bibfnamefont {M.}~\bibnamefont {Takada}},\ }\href@noop {}
  {\  (\bibinfo {year} {2014})},\ \Eprint {http://arxiv.org/abs/1401.0385}
  {arXiv:1401.0385 [astro-ph.CO]} \BibitemShut {NoStop}%
%%CITATION = ARXIV:1401.0385;%%
\bibitem [{\citenamefont {Hu}(2001{\natexlab{a}})}]{Hu:2001tn}%
  \BibitemOpen
  \bibfield  {author} {\bibinfo {author} {\bibfnamefont {W.}~\bibnamefont
  {Hu}},\ }\href {\doibase 10.1086/323253} {\bibfield  {journal} {\bibinfo
  {journal} {Astrophys.J.}\ }\textbf {\bibinfo {volume} {557}},\ \bibinfo
  {pages} {L79} (\bibinfo {year} {2001}{\natexlab{a}})},\ \Eprint
  {http://arxiv.org/abs/astro-ph/0105424} {arXiv:astro-ph/0105424 [astro-ph]}
  \BibitemShut {NoStop}%
%%CITATION = ASTRO-PH/0105424;%%
\bibitem [{\citenamefont {Okamoto}\ and\ \citenamefont
  {Hu}(2002)}]{Okamoto:2002ik}%
  \BibitemOpen
  \bibfield  {author} {\bibinfo {author} {\bibfnamefont {T.}~\bibnamefont
  {Okamoto}}\ and\ \bibinfo {author} {\bibfnamefont {W.}~\bibnamefont {Hu}},\
  }\href {\doibase 10.1103/PhysRevD.66.063008} {\bibfield  {journal} {\bibinfo
  {journal} {Phys.Rev.}\ }\textbf {\bibinfo {volume} {D66}},\ \bibinfo {pages}
  {063008} (\bibinfo {year} {2002})},\ \Eprint
  {http://arxiv.org/abs/astro-ph/0206155} {arXiv:astro-ph/0206155 [astro-ph]}
  \BibitemShut {NoStop}%
%%CITATION = ASTRO-PH/0206155;%%
\bibitem [{\citenamefont {Hanson}\ \emph {et~al.}(2011)\citenamefont {Hanson},
  \citenamefont {Challinor}, \citenamefont {Efstathiou},\ and\ \citenamefont
  {Bielewicz}}]{Hanson:2010rp}%
  \BibitemOpen
  \bibfield  {author} {\bibinfo {author} {\bibfnamefont {D.}~\bibnamefont
  {Hanson}}, \bibinfo {author} {\bibfnamefont {A.}~\bibnamefont {Challinor}},
  \bibinfo {author} {\bibfnamefont {G.}~\bibnamefont {Efstathiou}}, \ and\
  \bibinfo {author} {\bibfnamefont {P.}~\bibnamefont {Bielewicz}},\ }\href
  {\doibase 10.1103/PhysRevD.83.043005} {\bibfield  {journal} {\bibinfo
  {journal} {Phys.Rev.}\ }\textbf {\bibinfo {volume} {D83}},\ \bibinfo {pages}
  {043005} (\bibinfo {year} {2011})},\ \Eprint {http://arxiv.org/abs/1008.4403}
  {arXiv:1008.4403 [astro-ph.CO]} \BibitemShut {NoStop}%
\bibitem [{\citenamefont {Lewis}\ \emph {et~al.}(2011)\citenamefont {Lewis},
  \citenamefont {Challinor},\ and\ \citenamefont {Hanson}}]{Lewis:2011fk}%
  \BibitemOpen
  \bibfield  {author} {\bibinfo {author} {\bibfnamefont {A.}~\bibnamefont
  {Lewis}}, \bibinfo {author} {\bibfnamefont {A.}~\bibnamefont {Challinor}}, \
  and\ \bibinfo {author} {\bibfnamefont {D.}~\bibnamefont {Hanson}},\ }\href
  {\doibase 10.1088/1475-7516/2011/03/018} {\bibfield  {journal} {\bibinfo
  {journal} {JCAP}\ }\textbf {\bibinfo {volume} {1103}},\ \bibinfo {pages}
  {018} (\bibinfo {year} {2011})},\ \Eprint {http://arxiv.org/abs/1101.2234}
  {arXiv:1101.2234 [astro-ph.CO]} \BibitemShut {NoStop}%
%%CITATION = ARXIV:1101.2234;%%
\bibitem [{Note1()}]{Note1}%
  \BibitemOpen
  \bibinfo {note} {\protect \href
  {http://www.sciops.esa.int/wikiSI/planckpla}{Planck Explanatory
  Supplement}}\BibitemShut {NoStop}%
\bibitem [{\citenamefont {Hu}(2001{\natexlab{b}})}]{Hu:2001fa}%
  \BibitemOpen
  \bibfield  {author} {\bibinfo {author} {\bibfnamefont {W.}~\bibnamefont
  {Hu}},\ }\href {\doibase 10.1103/PhysRevD.64.083005} {\bibfield  {journal}
  {\bibinfo  {journal} {Phys.Rev.}\ }\textbf {\bibinfo {volume} {D64}},\
  \bibinfo {pages} {083005} (\bibinfo {year} {2001}{\natexlab{b}})},\ \Eprint
  {http://arxiv.org/abs/astro-ph/0105117} {arXiv:astro-ph/0105117 [astro-ph]}
  \BibitemShut {NoStop}%
%%CITATION = ASTRO-PH/0105117;%%
\bibitem [{\citenamefont {Hu}(2000)}]{Hu:2000ee}%
  \BibitemOpen
  \bibfield  {author} {\bibinfo {author} {\bibfnamefont {W.}~\bibnamefont
  {Hu}},\ }\href {\doibase 10.1103/PhysRevD.62.043007} {\bibfield  {journal}
  {\bibinfo  {journal} {Phys.Rev.}\ }\textbf {\bibinfo {volume} {D62}},\
  \bibinfo {pages} {043007} (\bibinfo {year} {2000})},\ \Eprint
  {http://arxiv.org/abs/astro-ph/0001303} {arXiv:astro-ph/0001303 [astro-ph]}
  \BibitemShut {NoStop}%
%%CITATION = ASTRO-PH/0001303;%%
\bibitem [{\citenamefont {{Okamoto}}\ and\ \citenamefont
  {{Hu}}(2003)}]{Okamoto03}%
  \BibitemOpen
  \bibfield  {author} {\bibinfo {author} {\bibfnamefont {T.}~\bibnamefont
  {{Okamoto}}}\ and\ \bibinfo {author} {\bibfnamefont {W.}~\bibnamefont
  {{Hu}}},\ }\href {\doibase 10.1103/PhysRevD.67.083002} {\bibfield  {journal}
  {\bibinfo  {journal} {Phys. Rev. D}\ }\textbf {\bibinfo {volume} {67}},\
  \bibinfo {pages} {083002} (\bibinfo {year} {2003})},\ \Eprint
  {http://arxiv.org/abs/arXiv:astro-ph/0301031} {arXiv:astro-ph/0301031}
  \BibitemShut {NoStop}%
\bibitem [{\citenamefont {Miranda}\ and\ \citenamefont
  {Hu}(2013)}]{Miranda:2013wxa}%
  \BibitemOpen
  \bibfield  {author} {\bibinfo {author} {\bibfnamefont {V.}~\bibnamefont
  {Miranda}}\ and\ \bibinfo {author} {\bibfnamefont {W.}~\bibnamefont {Hu}},\
  }\href@noop {} {\  (\bibinfo {year} {2013})},\ \Eprint
  {http://arxiv.org/abs/1312.0946} {arXiv:1312.0946 [astro-ph.CO]} \BibitemShut
  {NoStop}%
%%CITATION = ARXIV:1312.0946;%%
\end{thebibliography}%

 \end{document}